\documentclass[preprint]{vgtc}

\preprinttext{\small Accepted for presentation at the
  \href{https://visxgenai.github.io/}{VISxGenAI Workshop}, IEEE VIS 2026.}

\graphicspath{{figures/}}

\usepackage{times}

\usepackage{mathptmx}
\usepackage{balance}
\usepackage{dblfloatfix}
\usepackage{placeins}
\DeclareRobustCommand{\panel}[1]{\csname panel#1\endcsname}

\DeclareRobustCommand{\activityicon}{%
  \raisebox{-0.22em}{\includegraphics[height=1.25em]{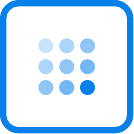}}%
}

\DeclareRobustCommand{\revealicon}{%
  \raisebox{-0.22em}{\includegraphics[height=1.25em]{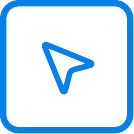}}%
}

\onlineid{1219}
\vgtccategory{Research}
\vgtcinsertpkg
\usepackage{hyperref} % Explicit declaration for the arXiv source scanner.

\usepackage{orcidlink}
\DeclareRobustCommand{\authororcid}[2]{#1\,\orcidlink{#2}}

\title{Point, Revise, Review: Grounded Agentic Analysis\\
in Reactive Notebooks with marimo-lens}
\author{%
  \authororcid{Péter Ferenc Gyarmati}{0009-0006-6122-2709},
  \authororcid{Trevor Manz}{0000-0001-7694-5164},
  \authororcid{Dominik Moritz}{0000-0002-3110-1053},
  \authororcid{Mennatallah El-Assady}{0000-0001-8526-2613}
}

\authorfooter{
  \item
  Péter Ferenc Gyarmati and Mennatallah El-Assady are\newline
  with ETH Zürich.
  \item
  Trevor Manz is with the marimo team at CoreWeave, Inc.
  \item
  Dominik Moritz is with Carnegie Mellon University.
}

\makeatletter
\renewcommand{\copyrightspace}{%
  \ifreviewelse{}{%
    \renewcommand{\thefootnote}{}%
    \footnotetext[0]{%
      \begin{flushleft}
      \vskip -6pt
      \begin{list}{\textbullet}{%
        \setlength{\partopsep}{0pt}%
        \setlength{\topsep}{0pt}%
        \setlength{\itemsep}{-2pt}%
        \setlength{\itemindent}{-4pt}%
        \setlength{\leftmargin}{12pt}}
        \authorfootertext
      \end{list}
      \vskip 4pt
      \end{flushleft}}%
    \renewcommand{\thefootnote}{\arabic{footnote}}%
  }%
}
\makeatother

\teaser{
  \centering
  \makebox[\linewidth][c]{%
    \includegraphics[width=\dimexpr\linewidth+.66in\relax]{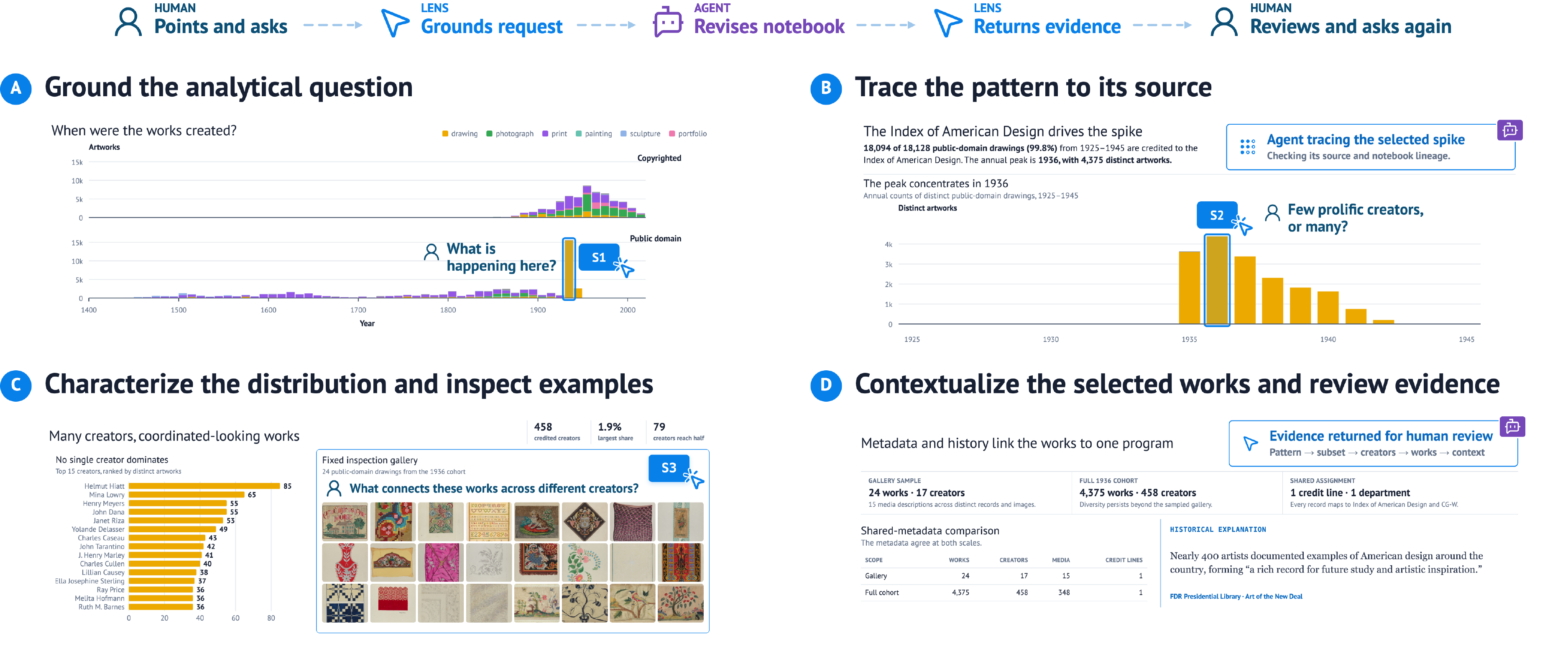}%
  }
  \caption{\textbf{A grounded human-agent analysis with marimo-lens.} The human marks a notebook result and asks a situated question~\panel{A}. Lens connects that request to the underlying computation, and the agent returns a focused result~\panel{B}. A subsequent request characterizes the cohort, while selected works provide a surface for further inspection~\panel{C}. Lens then returns visual, metadata, and historical evidence to the notebook for human review~\panel{D}. Each returned result can become the starting point for the next analytical cycle.}
  \label{fig:overview}
}

\abstract{In a computational notebook, a human can point to a rendered result and ask about ``this,'' while an agent acts through cells, dependencies, and runtime state. When what the human sees and what the agent operates on are disconnected, the human must describe what they mean and trace what the agent did in prose that strips away situated visual and computational context. We present \textit{marimo-lens}, an extension to the reactive Python notebook marimo for grounded human-agent analysis. Lens connects a human's marked output and note to its producing cell and relevant contributing computation, giving the agent computational context for the request. It surfaces agent activity, returns selected results to the notebook, and preserves the initiating selection for human review and reopening. We illustrate the lifecycle through an exploratory human-agent analysis of a real-world open dataset, showing how visually situated questions lead to computational inspection, notebook action, and returned evidence for human review.}


\hypersetup{
  pdftitle={Point, Revise, Review: Grounded Agentic Analysis in Reactive Notebooks with marimo-lens},
  pdfauthor={Péter Ferenc Gyarmati, Trevor Manz, Dominik Moritz, Mennatallah El-Assady},
  pdfsubject={Accepted for presentation at the VISxGenAI Workshop, IEEE VIS 2026},
  pdfkeywords={human-agent collaboration, reactive notebooks, visual analytics, visual grounding}
}

\begin{document}

\firstsection{Introduction}
\maketitle

\begin{figure*}[!t]
  \centering
  \includegraphics[width=\textwidth]{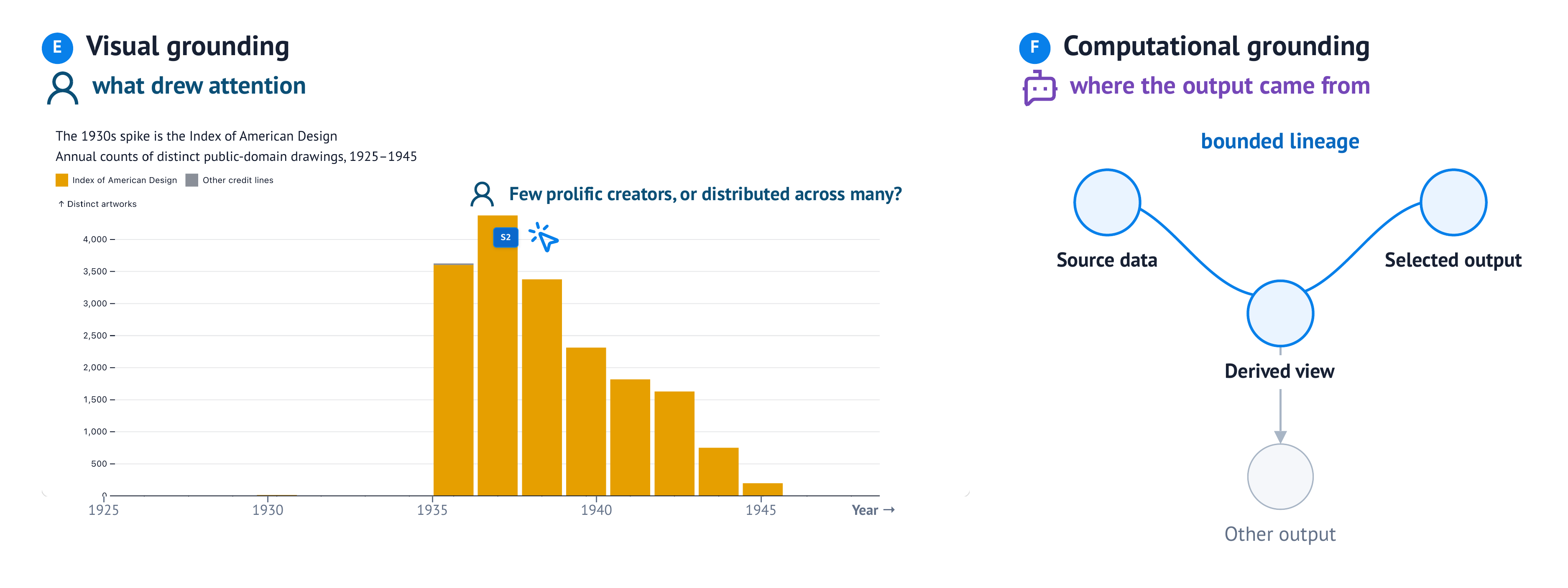}
  \caption{\textbf{Connecting a marked result to its computation.} The rendered output, mark, and question preserve what drew the analyst's attention~\panel{E}. Lens connects that reference to the producing cell and a bounded view of the computation that contributes to the selected output~\panel{F}. Pair~\cite{marimoCollaborateAgentsUsing2026} uses this context as a starting point for working in the live notebook, while marimo~\cite{marimoMarimoReactivePython2023} reruns cells affected by the agent's changes.}
  \label{fig:grounding}
\end{figure*}

\begin{figure*}[t]
  \centering
  \includegraphics[width=\textwidth]{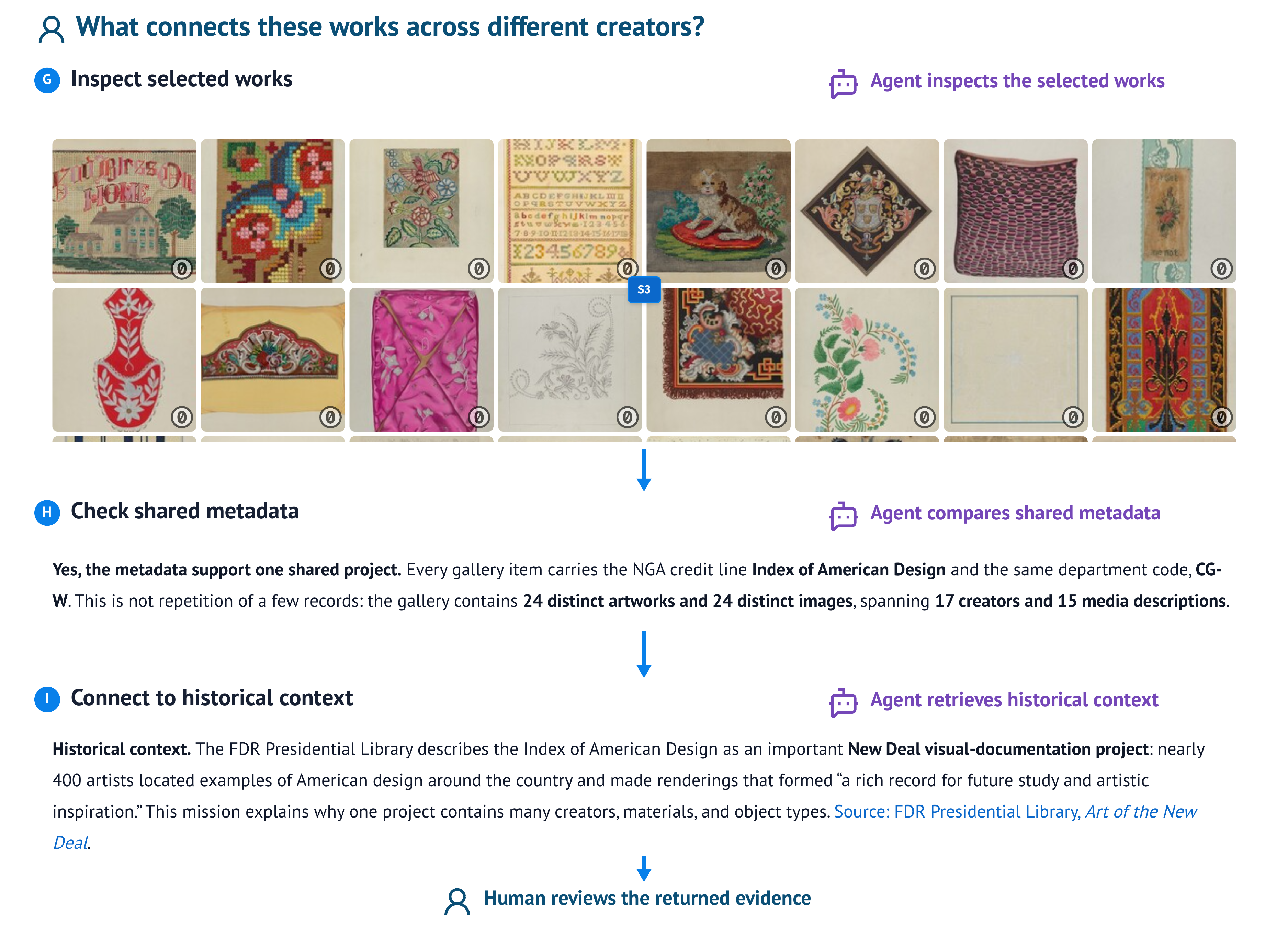}
  \caption{\textbf{Characterizing a pattern noticed by the analyst.} Lens returns three forms of evidence for human review: selected works~\panel{G}, metadata shared across the gallery and broader cohort~\panel{H}, and historical context about the Index of American Design~\panel{I}. Together, these views help the analyst interpret what connects the selected works.}
  \label{fig:evidence}
\end{figure*}

As data analysis becomes increasingly agentic~\cite{dhanoa_agenticVis_2025}, computational notebooks must support interaction between analysts who reason through visible results and agents that operate on code, dependencies, and runtime state. In \autoref{fig:overview}\,\panel{A}, a human marks a concentration in a rendered output and asks, ``What is happening here?'' This combination of pointing and language is a form of deixis: the mark identifies what drew the analyst's attention, while the accompanying question gives that reference meaning. Cursor gestures become interpretable when preserved with language~\cite{hanDeixisCenteredApproachDocumenting2025a}, and computational graphics can reify ephemeral pointing so it remains associated with an image~\cite{hillDeixisFutureVisualization1991}. Because reference is collaborative and repairable~\cite{clarkReferringCollaborativeProcess1986}, preserving the mark together with the question creates a situated analytical request. For an agent to inspect or revise the analysis, that request must also connect to the notebook computation that produced the marked output.

This requirement exposes a representational gap in agent-assisted notebook analysis. Analysts refer through rendered output, spatial attention, language, and judgment, while agents act through cells, code, dependencies, live values, and execution. Notebook systems vary in how they represent and execute these artifacts~\cite{lauDesignSpaceComputational2020}. For assistants working in these notebooks, McNutt et al. identify multimodal context, specification, refinement, and verification as central design concerns~\cite{mcnuttDesignAIpoweredCode2023a}. Grounding a situated request therefore requires both a visible reference to what the analyst means and a computational route to the notebook state an agent can inspect and revise.

Prior systems connect parts of this path. DirectGPT binds prompt actions to visible objects~\cite{massonDirectGPTDirectManipulation2024}, DirectVis maps D3 elements to code through shared state~\cite{parkDirectVisEditingCodeBased2026a}, and B2 uses data queries to bridge notebook code with interactive output~\cite{wuB22020}. NoteEx lets analysts curate notebook context~\cite{payandehNoteExInteractiveVisual2025}, while VACP exposes structured state and executable interactions to visual-analytics agents~\cite{stahleVACPVisualAnalytics2026}. Together, these systems establish direct reference, code-view correspondence, curated context, and executable action. Our focus is an end-to-end request lifecycle that starts from a mark on a rendered notebook output, connects that visual reference and its accompanying note to the producing cell and relevant contributing computation, and returns agent-produced evidence to the notebook surface for human review.

This lifecycle also defines how initiative is distributed between human and agent. Agentic visualization treats role allocation and human control as design concerns~\cite{dhanoa_agenticVis_2025}, while mixed-initiative visual analytics spans varied levels of automation and raises questions about human agency~\cite{monadjemiScopingReviewMixed2026}. Lens is designed around an allocation in which the human points, asks, interprets, and judges returned evidence, while the agent operates on the notebook within each grounded request. Following Holter and El-Assady's distinction among agency, interaction, and adaptation~\cite{holterDeconstructingHumanAICollaboration2024a}, we separate human direction and judgment from agent action.

Supporting this allocation requires the notebook to connect three parts of the interaction: the human's visual reference, the agent's computational actions, and the evidence returned for human review. We therefore ask: \emph{How can a reactive notebook bind a human-marked result to computation an agent can inspect and revise, then return evidence to a review point the human can revisit?}

We present \emph{marimo-lens} (Lens),\footnote{Source code: \url{https://github.com/marimo-team/marimo-lens}.} an open-source tool that augments marimo, a reactive Python notebook~\cite{marimoMarimoReactivePython2023}, and coordinates with its embedded notebook agent, \emph{marimo-pair} (Pair)~\cite{marimoCollaborateAgentsUsing2026}. Our contribution is an implemented request lifecycle that connects a human mark to the computation that produced the marked output, coordinates agent work in the live notebook, and returns results to the notebook surface for human review. Lens uses the output cell as the grounding unit, gives the agent a bounded view of the computation contributing to that output, and preserves the initiating selection so the human can return to it later.

\begin{figure*}[!t]
  \centering
  \includegraphics[width=\textwidth]{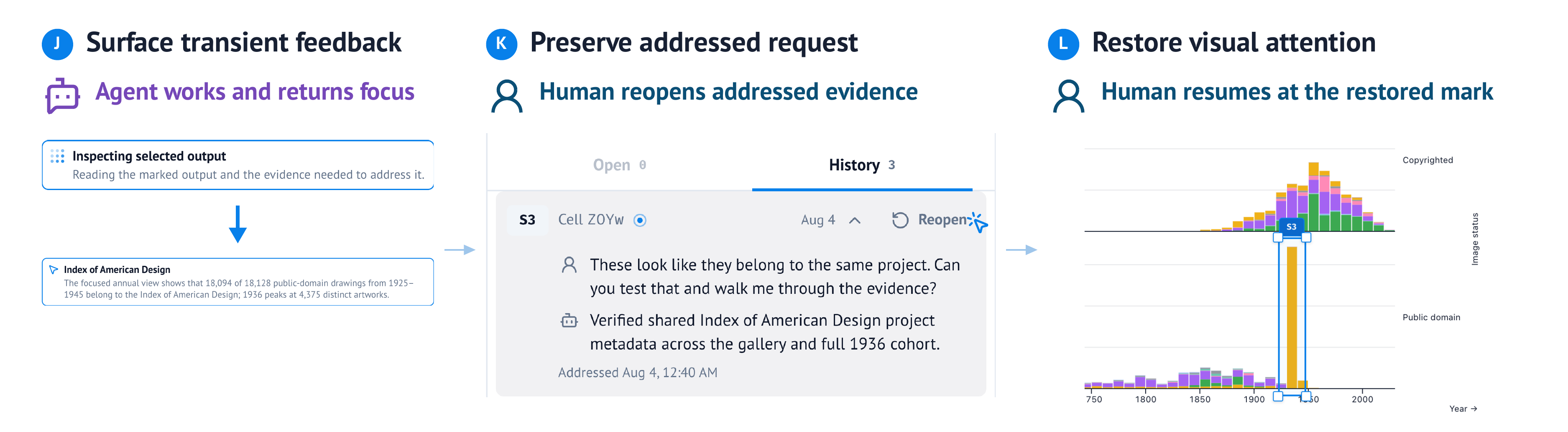}
  \caption{\textbf{From agent action to human review and reopening.} Lens keeps the agent's activity and returned result visible~\panel{J}, preserves the initiating request when the agent finishes addressing it~\panel{K}, and lets the analyst reopen the mark and note on the current output~\panel{L}. This creates a review point between cycles of agent action.}
  \label{fig:lifecycle}
\end{figure*}

We illustrate this lifecycle through three grounded requests in a human-agent exploratory analysis of an open dataset from the National Gallery of Art~\cite{nationalgalleryofartNationalGalleryArt2021}. The accompanying executable notebook and interface recording document the implemented workflow. \autoref{fig:grounding} explains how a marked output is connected to its computation, \autoref{fig:evidence} shows the analytical views returned for review, and \autoref{fig:lifecycle} shows how a request moves from agent activity to reopening.

\section{Connecting Marked Outputs to Computation}
\label{sec:grounding}

We call the marked selection together with its producing cell and bounded computational context a \emph{grounded request}. \autoref{fig:grounding} shows how one grounded request spans Lens, Pair, and marimo. Lens records the mark, note, producing cell, and selection revision, then uses marimo's dependency graph to gather a bounded view of the computation that contributes to the selected output. Pair uses this context as a starting point for inspecting and revising the live notebook~\cite{marimoCollaborateAgentsUsing2026}. As the agent makes changes, marimo updates the dependency graph and reruns affected downstream cells~\cite{marimoReactivityDocumentation2026}. Lens manages how that work is surfaced and revisited by the analyst, including activity feedback, returned results, and reopening.

Lens grounds the human's request in both the rendered result and its producing computation. When Pair requests context from the active notebook kernel, Lens returns coordinated selection references, optional image evidence, and bounded notebook text. Each selection reference includes the human note, output-cell identifier, normalized point or rectangle geometry, capture and cell-availability status, and optional DOM metadata such as role, accessible label, visible text, and bounds. When capture succeeds, Lens provides an annotated PNG of the rendered output, with the selected point or region marked directly in image space. This follows visual prompting approaches that overlay marks on images to direct multimodal models to specific regions~\cite{caiViPLLaVAMakingLarge2024}. For large or scrolled outputs, the image combines an overview with a focused detail. The text projection describes producing and upstream cells through their source, language, defined and referenced names, direct upstream identifiers, and relevant native-control state. Together, these representations connect the marked visual referent to computation Pair can inspect in the live notebook.

Lens derives the bounded notebook context from the producing cells of all open selections. It traverses direct parent edges breadth-first, prioritizing selected outputs and nearby ancestors, and retains at most 64 cells while allocating a shared 24,000-character source budget in the same order. Retained cells are presented topologically so dependencies precede their consumers, and Lens reports omitted cells and source truncation when these limits are reached. These bounds limit unrelated notebook state and avoid relying on arbitrarily long contexts, which language models can use unreliably~\cite{liuLostMiddleHow2024}. Unlike NoteEx's human-curated context~\cite{payandehNoteExInteractiveVisual2025}, this context is derived automatically from the marked outputs and the notebook's current dependency graph. \autoref{fig:grounding}\,\panel{F} illustrates this bounded view of the computation behind the selected output.

This bounded context gives Pair a concrete starting point for working in the live notebook while leaving it free to inspect and revise the broader notebook. From there, the agent can inspect values and outputs, read or edit existing cells, add new cells, and execute code as needed. These capabilities support the specification and verification activities that McNutt et al. identify for notebook assistants~\cite{mcnuttDesignAIpoweredCode2023a}, and echo Flowco's use of dataflow structure for mixed-initiative analysis authoring~\cite{freundFlowcoMixedInitiativeAuthoring2025}.

As the agent works, Lens shows the agent's activity and results on the notebook surface so the human analyst can follow both the analytical process and the evidence it produces. The agent calls \activityicon\,\texttt{start\_activity()} to indicate the cell it is working on, \revealicon\,\texttt{reveal()} to bring a selected result into view, and \texttt{resolve()} when it has addressed the request. If the initiating selection is still current, \texttt{resolve()} stores an addressed receipt containing the request, its mark, cell identity, timestamps, and an optional summary. The analyst can later reopen the request, restoring the mark and note on the current output and continuing from that review point. Activity and reveal make the agent's current work visible, while resolve and reopen preserve a review point across analytical turns. Related systems use similar monitoring and inspection cues for agentic data analysis~\cite{WaitGPTMonitoringSteering2024,wuStepMINDVisualFramework2026}.

\section{Human-in-the-Loop Agentic Analysis with Lens}
\label{sec:walkthrough}

We illustrate Lens through an exploratory analysis of an open dataset from the National Gallery of Art~\cite{nationalgalleryofartNationalGalleryArt2021,nationalgalleryofartOpenDataCommit2026}. The analysis takes place in a single marimo notebook and follows three successive requests in which the human marks a result, asks a situated question, and reviews the evidence returned through Lens. The supplement provides the starting and resulting notebooks, a short interface recording, and the agent harness and configuration used for the paired analysis.

The first request starts from an overview in which the human notices a concentration of drawings with open-access images. They mark this region and ask, ``What is happening here?'' Lens connects the marked output to the computation that produced it, and Pair returns a focused timeline. The result shows that 18,094 of 18,128 drawings from 1925 to 1945 are associated with the Index of American Design, peaking at 4,375 works in 1936.

The human then marks the 1936 peak and asks whether a small number of prolific creators account for the concentration. Pair computes the creator distribution for this cohort: the most frequent creator accounts for only 1.9\% of the works, and it takes 79 creators to account for half of them. The concentration is therefore not the product of a few prolific creators. Pair also returns a gallery of sampled works from the same cohort as a visual surface for further inspection.

From this gallery, the human notices related visual characteristics across works attributed to different creators and asks, ``What connects these works across different creators?'' \autoref{fig:evidence} brings together three forms of evidence: selected works~\panel{G}, metadata shared across the gallery and broader cohort~\panel{H}, and historical context~\panel{I}. The historical account describes the Index of American Design as a New Deal program in which nearly 400 artists documented examples of American design~\cite{franklind.rooseveltpresidentiallibraryandmuseumArtNewDeal}. Lens returns these views to the notebook surface so the human can inspect how the emerging interpretation is supported by visual, computational, and historical evidence.

Across the three requests, the marked output serves as a visible reference for the human and an entry point to the underlying computation for the agent. Lens keeps the agent's work visible on the notebook surface and returns each result to a review point where the analyst can decide how the analysis should continue.

\section{Human Review and Reopening}
\label{sec:lens-history}

The walkthrough in \autoref{sec:walkthrough} shows how each returned result creates a new point for human interpretation and direction. Keeping the analyst involved between these cycles lets them inspect and question intermediate results, and redirect the analysis before it continues~\cite{amershiGuidelinesHumanAIInteraction2019}. Prior work has shown the value of preserving analytical state for revisitation, including graphical histories~\cite{heerGraphicalHistoriesVisualization2008b}, notebook versions~\cite{eckeltLoopsLeveragingProvenance2025a}, and reusable interaction histories~\cite{gadhavePersistPersistentReusable2024a}. Lens carries this idea into human-agent analysis by returning each addressed request to a review point the analyst can revisit.

As shown in \autoref{fig:lifecycle}, activity \activityicon{} and reveal \revealicon{} keep the agent's work visible while it is in progress~\panel{J}. When the agent resolves the request, Lens preserves a compact record of the initiating mark, note, and producing cell~\panel{K}. This creates an explicit handoff from agent action back to the analyst, complementing prior work on preserving analytical context across collaborative handoffs~\cite{zhaoSupportingHandoffAsynchronous2018}. Resolution records that the agent has addressed the request, while the human remains responsible for judging the returned evidence.

The analyst can later reopen the request to restore its mark and note on the current output~\panel{L}. From there, they can inspect the result again, refine the mark or question when the agent interpreted the wrong feature or returned incomplete evidence, or simply continue the analysis. This ability to revisit and repair a situated request follows the collaborative nature of reference~\cite{clarkReferringCollaborativeProcess1986} and related work on structured grounding for collaborative revision~\cite{doGroundingStructureExploring2024}. Lens therefore preserves a concrete point at which the analyst can review, repair, or redirect the analysis before the next cycle of agent action.

\section{Discussion and Limitations}
\label{sec:discussion}

Lens suggests a broader design principle for agentic visual analysis: human attention, agent action, and human review should remain connected throughout the analytical process. A visual reference gives the agent a situated starting point, while its connection to notebook computation exposes the state the agent can inspect and revise. Returning the agent's work to the same notebook surface then gives the analyst a concrete place to inspect, question, and redirect what happened. This complements prior work on exposing structured analytical state to agents~\cite{stahleVACPVisualAnalytics2026} and on using visual artifacts as anchors for human-LLM analysis~\cite{elshehalyDesigningCollaborationVisualization2025a}. More broadly, Lens treats grounding as a continuing relationship between what the human attends to, what the agent can act on, and what the human can subsequently review.

Several limitations remain. Lens connects a marked region to its producing cell and relevant contributing computation, but the agent still has to interpret what the human means within that region. The computational context is deliberately bounded, so relevant information elsewhere in a large notebook may be omitted. The notebook can also change between a request and a later reopening, causing an old mark to refer to a different visual feature. Finally, our walkthrough illustrates the lifecycle in one notebook with one agent configuration and one exploratory analysis. It demonstrates how the interaction can work, while its effects on analytical accuracy, efficiency, and human understanding remain to be established.

Studies of visual grounding can test whether marked outputs and notes help agents identify what analysts mean~\cite{niuChartREGBenchmarkingImproving2026a}. Studies of computational grounding can test whether connecting those references to notebook computation helps agents inspect and revise the intended analysis~\cite{zhaoChartEditHowFar2025,tangChartsCodeHierarchical2026}. Multi-turn studies can examine whether visible agent activity, returned results, and reopening help analysts understand, correct, and continue an analysis over time~\cite{kapadnisChartEditBenchEvaluatingGrounded2026}. Together, these questions would test whether the lifecycle demonstrated by Lens translates into more effective human-agent analysis across notebooks, agents, and tasks.

\section{Conclusion}
\label{sec:conclusion}

marimo-lens connects what analysts attend to with what agents can inspect and change, then returns agent work to a concrete point for human review. Across the walkthrough, the analyst directs attention, interprets evidence, and decides how the analysis should proceed, while the agent carries out computational inspection and revision in the notebook. This interaction model preserves analyst agency as data analysis becomes increasingly agentic: agents extend the analyst's computational reach, while the human retains control over analytical direction and judgment. By grounding agent action in visible results and reconnecting it to human review, Lens provides a concrete model for human-in-the-loop agentic analysis in reactive notebooks.

\acknowledgments{We thank the marimo team for their continued feedback and support throughout the research and development of marimo-lens.}

\clearpage
\begingroup
\renewcommand{\baselinestretch}{0.9}\selectfont
\setlength{\bibspacing}{1.75pt}
\bibliographystyle{abbrv-doi-hyperref-narrow}
\bibliography{references}

\begin{thebibliography}{10}
\renewcommand*{\sfdefault}{PTSansNarrow-TLF}

\bibitem{amershiGuidelinesHumanAIInteraction2019}
\href{https://doi.org/10.1145/3290605.3300233}{S.~Amershi, D.~Weld,
  M.~Vorvoreanu, A.~Fourney, B.~Nushi, P.~Collisson, J.~Suh, S.~Iqbal, P.~N.
  Bennett, K.~Inkpen, J.~Teevan, R.~Kikin-Gil, and E.~Horvitz}.
\newblock \href{https://doi.org/10.1145/3290605.3300233}{{Guidelines for
  {Human}-{AI} {Interaction}}}.
\newblock \href{https://doi.org/10.1145/3290605.3300233}{In {\em Proceedings of
  the 2019 {CHI} {Conference} on {Human} {Factors} in {Computing} {Systems}}},
  \href{https://doi.org/10.1145/3290605.3300233}{May 2019}.
  \href{https://doi.org/10.1145/3290605.3300233}
{doi: \textsf{%
10\hspace{.1pt}\discretionary{.}{%
}{.}\hspace{.4pt}1145\discretionary{/}{%
}{/}3290605\hspace{.1pt}\discretionary{.}{%
}{.}\hspace{.4pt}3300233}}


\bibitem{caiViPLLaVAMakingLarge2024}
\href{https://doi.org/10.1109/CVPR52733.2024.01227}{M.~Cai, H.~Liu, S.~K.
  Mustikovela, G.~P. Meyer, Y.~Chai, D.~Park, and Y.~J. Lee}.
\newblock \href{https://doi.org/10.1109/CVPR52733.2024.01227}{{{ViP}-{LLaVA}:
  {Making} {Large} {Multimodal} {Models} {Understand} {Arbitrary} {Visual}
  {Prompts}}}.
\newblock \href{https://doi.org/10.1109/CVPR52733.2024.01227}{pp.
  12914--12923}. \href{https://doi.org/10.1109/CVPR52733.2024.01227}{IEEE
  Computer Society}, \href{https://doi.org/10.1109/CVPR52733.2024.01227}{June
  2024}. \href{https://doi.org/10.1109/CVPR52733.2024.01227}
{doi: \textsf{%
10\hspace{.1pt}\discretionary{.}{%
}{.}\hspace{.4pt}1109\discretionary{/}{%
}{/}CVPR52733\hspace{.1pt}\discretionary{.}{%
}{.}\hspace{.4pt}2024\hspace{.1pt}\discretionary{.}{%
}{.}\hspace{.4pt}01227}}


\bibitem{clarkReferringCollaborativeProcess1986}
\href{https://doi.org/10.1016/0010-0277(86)90010-7}{H.~H. Clark and
  D.~Wilkes-Gibbs}.
\newblock \href{https://doi.org/10.1016/0010-0277(86)90010-7}{{Referring as a
  collaborative process}}.
\newblock \href{https://doi.org/10.1016/0010-0277(86)90010-7}{{\em Cognition}},
  \href{https://doi.org/10.1016/0010-0277(86)90010-7}{22(1):1--39},
  \href{https://doi.org/10.1016/0010-0277(86)90010-7}{Feb. 1986}.
  \href{https://doi.org/10.1016/0010-0277(86)90010-7}
{doi: \textsf{%
10\hspace{.1pt}\discretionary{.}{%
}{.}\hspace{.4pt}1016\discretionary{/}{%
}{/}0010\discretionary{%
}{-}{-}0277\discretionary{%
}{(}{(}86\discretionary{)}{%
}{)}90010\discretionary{%
}{-}{-}7}}


\bibitem{dhanoa_agenticVis_2025}
\href{https://doi.org/10.1109/MCG.2025.3607741}{V.~Dhanoa, A.~Wolter, G.~M.
  Le\'{o}n, H.-J. Schulz, and N.~Elmqvist}.
\newblock \href{https://doi.org/10.1109/MCG.2025.3607741}{{ {Agentic
  {Visualization}: {Extracting} {Agent}-{Based} {Design} {Patterns} {From}
  {Visualization} {Systems}} }}.
\newblock \href{https://doi.org/10.1109/MCG.2025.3607741}{{\em IEEE Computer
  Graphics and Applications}},
  \href{https://doi.org/10.1109/MCG.2025.3607741}{45(6):89--100},
  \href{https://doi.org/10.1109/MCG.2025.3607741}{Nov. 2025}.
  \href{https://doi.org/10.1109/MCG.2025.3607741}
{doi: \textsf{%
10\hspace{.1pt}\discretionary{.}{%
}{.}\hspace{.4pt}1109\discretionary{/}{%
}{/}MCG\hspace{.1pt}\discretionary{.}{%
}{.}\hspace{.4pt}2025\hspace{.1pt}\discretionary{.}{%
}{.}\hspace{.4pt}3607741}}


\bibitem{doGroundingStructureExploring2024}
\href{https://doi.org/10.1145/3686902}{H.~J. Do, M.~Brachman, C.~Dugan, J.~M.
  Johnson, J.~Lauer, P.~Rai, and Q.~Pan}.
\newblock \href{https://doi.org/10.1145/3686902}{{ {Grounding with {Structure}:
  {Exploring} {Design} {Variations} of {Grounded} {Human}-{AI} {Collaboration}
  in a {Natural} {Language} {Interface}} }}.
\newblock \href{https://doi.org/10.1145/3686902}{{\em Proceedings of the ACM on
  Human-Computer Interaction}},
  \href{https://doi.org/10.1145/3686902}{8(CSCW2):363:1--363:27},
  \href{https://doi.org/10.1145/3686902}{Nov. 2024}.
  \href{https://doi.org/10.1145/3686902}
{doi: \textsf{%
10\hspace{.1pt}\discretionary{.}{%
}{.}\hspace{.4pt}1145\discretionary{/}{%
}{/}3686902}}


\bibitem{eckeltLoopsLeveragingProvenance2025a}
\href{https://doi.org/10.1109/tvcg.2024.3456186}{K.~Eckelt, K.~Gadhave, A.~Lex,
  and M.~Streit}.
\newblock \href{https://doi.org/10.1109/tvcg.2024.3456186}{{ {Loops:
  {Leveraging} {Provenance} and {Visualization} to {Support} {Exploratory}
  {Data} {Analysis} in {Notebooks}} }}.
\newblock \href{https://doi.org/10.1109/tvcg.2024.3456186}{{\em IEEE
  Transactions on Visualization and Computer Graphics}},
  \href{https://doi.org/10.1109/tvcg.2024.3456186}{Jan. 2025}.
  \href{https://doi.org/10.1109/tvcg.2024.3456186}
{doi: \textsf{%
10\hspace{.1pt}\discretionary{.}{%
}{.}\hspace{.4pt}1109\discretionary{/}{%
}{/}tvcg\hspace{.1pt}\discretionary{.}{%
}{.}\hspace{.4pt}2024\hspace{.1pt}\discretionary{.}{%
}{.}\hspace{.4pt}3456186}}


\bibitem{elshehalyDesigningCollaborationVisualization2025a}
\href{https://doi.org/10.1109/MCG.2025.3583451}{M.~Elshehaly, R.~Jianu,
  A.~Slingsby, G.~Andrienko, and N.~Andrienko}.
\newblock \href{https://doi.org/10.1109/MCG.2025.3583451}{{ {Designing for
  {Collaboration}: {Visualization} to {Enable} {Human}–{LLM} {Analytical}
  {Partnership}} }}.
\newblock \href{https://doi.org/10.1109/MCG.2025.3583451}{{\em IEEE Computer
  Graphics and Applications}},
  \href{https://doi.org/10.1109/MCG.2025.3583451}{45(5):107--116},
  \href{https://doi.org/10.1109/MCG.2025.3583451}{Sept. 2025}.
  \href{https://doi.org/10.1109/MCG.2025.3583451}
{doi: \textsf{%
10\hspace{.1pt}\discretionary{.}{%
}{.}\hspace{.4pt}1109\discretionary{/}{%
}{/}MCG\hspace{.1pt}\discretionary{.}{%
}{.}\hspace{.4pt}2025\hspace{.1pt}\discretionary{.}{%
}{.}\hspace{.4pt}3583451}}


\bibitem{franklind.rooseveltpresidentiallibraryandmuseumArtNewDeal}
\href{https://www.fdrlibrary.org/art-detail}{{Franklin D. Roosevelt
  Presidential Library and Museum}}.
\newblock \href{https://www.fdrlibrary.org/art-detail}{{Art of the {New}
  {Deal}: {The} {Index} of {American} {Design}}}.

\bibitem{freundFlowcoMixedInitiativeAuthoring2025}
\href{https://doi.org/10.1145/3746059.3747636}{S.~N. Freund, B.~Simon, E.~D.
  Berger, and E.~Jun}.
\newblock \href{https://doi.org/10.1145/3746059.3747636}{{ {Flowco:
  {Mixed}-{Initiative} {Authoring} of {Reliable} {End}-to-{End} {Data}
  {Analyses} via {Dataflow} {Graphs} and {LLMs}} }}.
\newblock \href{https://doi.org/10.1145/3746059.3747636}{In {\em Proceedings of
  the 38th {Annual} {ACM} {Symposium} on {User} {Interface} {Software} and
  {Technology}}}, \href{https://doi.org/10.1145/3746059.3747636}{{UIST} '25},
  \href{https://doi.org/10.1145/3746059.3747636}{pp. 1--20}.
  \href{https://doi.org/10.1145/3746059.3747636}{Association for Computing
  Machinery}, \href{https://doi.org/10.1145/3746059.3747636}{New York, NY,
  USA}, \href{https://doi.org/10.1145/3746059.3747636}{Sept. 2025}.
  \href{https://doi.org/10.1145/3746059.3747636}
{doi: \textsf{%
10\hspace{.1pt}\discretionary{.}{%
}{.}\hspace{.4pt}1145\discretionary{/}{%
}{/}3746059\hspace{.1pt}\discretionary{.}{%
}{.}\hspace{.4pt}3747636}}


\bibitem{gadhavePersistPersistentReusable2024a}
\href{https://doi.org/10.1111/cgf.15092}{K.~Gadhave, Z.~Cutler, and A.~Lex}.
\newblock \href{https://doi.org/10.1111/cgf.15092}{{ {Persist: {Persistent} and
  {Reusable} {Interactions} in {Computational} {Notebooks}} }}.
\newblock \href{https://doi.org/10.1111/cgf.15092}{{\em Computer Graphics
  Forum}}, \href{https://doi.org/10.1111/cgf.15092}{43(3)},
  \href{https://doi.org/10.1111/cgf.15092}{June 2024}.
  \href{https://doi.org/10.1111/cgf.15092}
{doi: \textsf{%
10\hspace{.1pt}\discretionary{.}{%
}{.}\hspace{.4pt}1111\discretionary{/}{%
}{/}cgf\hspace{.1pt}\discretionary{.}{%
}{.}\hspace{.4pt}15092}}


\bibitem{hanDeixisCenteredApproachDocumenting2025a}
\href{https://doi.org/10.1109/TVCG.2024.3456351}{C.~Han and K.~E. Isaacs}.
\newblock \href{https://doi.org/10.1109/TVCG.2024.3456351}{{ {A
  {Deixis}-{Centered} {Approach} for {Documenting} {Remote} {Synchronous}
  {Communication} {Around} {Data} {Visualizations}} }}.
\newblock \href{https://doi.org/10.1109/TVCG.2024.3456351}{{\em IEEE
  Transactions on Visualization and Computer Graphics}},
  \href{https://doi.org/10.1109/TVCG.2024.3456351}{31(1):930--940},
  \href{https://doi.org/10.1109/TVCG.2024.3456351}{Jan. 2025}.
  \href{https://doi.org/10.1109/TVCG.2024.3456351}
{doi: \textsf{%
10\hspace{.1pt}\discretionary{.}{%
}{.}\hspace{.4pt}1109\discretionary{/}{%
}{/}TVCG\hspace{.1pt}\discretionary{.}{%
}{.}\hspace{.4pt}2024\hspace{.1pt}\discretionary{.}{%
}{.}\hspace{.4pt}3456351}}


\bibitem{heerGraphicalHistoriesVisualization2008b}
\href{https://doi.org/10.1109/TVCG.2008.137}{J.~Heer, J.~Mackinlay, C.~Stolte,
  and M.~Agrawala}.
\newblock \href{https://doi.org/10.1109/TVCG.2008.137}{{ {Graphical {Histories}
  for {Visualization}: {Supporting} {Analysis}, {Communication}, and
  {Evaluation}} }}.
\newblock \href{https://doi.org/10.1109/TVCG.2008.137}{{\em IEEE Transactions
  on Visualization and Computer Graphics}},
  \href{https://doi.org/10.1109/TVCG.2008.137}{14(6):1189--1196},
  \href{https://doi.org/10.1109/TVCG.2008.137}{Nov. 2008}.
  \href{https://doi.org/10.1109/TVCG.2008.137}
{doi: \textsf{%
10\hspace{.1pt}\discretionary{.}{%
}{.}\hspace{.4pt}1109\discretionary{/}{%
}{/}TVCG\hspace{.1pt}\discretionary{.}{%
}{.}\hspace{.4pt}2008\hspace{.1pt}\discretionary{.}{%
}{.}\hspace{.4pt}137}}


\bibitem{hillDeixisFutureVisualization1991}
\href{https://doi.org/10.1109/VISUAL.1991.175820}{W.~Hill and J.~Hollan}.
\newblock \href{https://doi.org/10.1109/VISUAL.1991.175820}{{Deixis and the
  future of visualization excellence}}.
\newblock \href{https://doi.org/10.1109/VISUAL.1991.175820}{In {\em Proceeding
  {Visualization} '91}}, \href{https://doi.org/10.1109/VISUAL.1991.175820}{pp.
  314--320}. \href{https://doi.org/10.1109/VISUAL.1991.175820}{IEEE Comput.
  Soc. Press}, \href{https://doi.org/10.1109/VISUAL.1991.175820}{San Diego, CA,
  USA}, \href{https://doi.org/10.1109/VISUAL.1991.175820}{Oct. 1991}.
  \href{https://doi.org/10.1109/VISUAL.1991.175820}
{doi: \textsf{%
10\hspace{.1pt}\discretionary{.}{%
}{.}\hspace{.4pt}1109\discretionary{/}{%
}{/}VISUAL\hspace{.1pt}\discretionary{.}{%
}{.}\hspace{.4pt}1991\hspace{.1pt}\discretionary{.}{%
}{.}\hspace{.4pt}175820}}


\bibitem{holterDeconstructingHumanAICollaboration2024a}
\href{https://doi.org/10.1111/cgf.15107}{S.~Holter and M.~El-Assady}.
\newblock \href{https://doi.org/10.1111/cgf.15107}{{ {Deconstructing
  {Human}-{AI} {Collaboration}: {Agency}, {Interaction}, and {Adaptation}} }}.
\newblock \href{https://doi.org/10.1111/cgf.15107}{{\em Computer Graphics
  Forum}}, \href{https://doi.org/10.1111/cgf.15107}{43(3):e15107},
  \href{https://doi.org/10.1111/cgf.15107}{2024}.
  \href{https://doi.org/10.1111/cgf.15107}
{doi: \textsf{%
10\hspace{.1pt}\discretionary{.}{%
}{.}\hspace{.4pt}1111\discretionary{/}{%
}{/}cgf\hspace{.1pt}\discretionary{.}{%
}{.}\hspace{.4pt}15107}}


\bibitem{kapadnisChartEditBenchEvaluatingGrounded2026}
\href{https://doi.org/10.48550/arXiv.2602.15758}{M.~N. Kapadnis, L.~Baghel,
  A.~Naik, and C.~Ros\'{e}}.
\newblock \href{https://doi.org/10.48550/arXiv.2602.15758}{{ {{ChartEditBench}:
  {Evaluating} {Grounded} {Multi}-{Turn} {Chart} {Editing} in {Multimodal}
  {Language} {Models}} }},
  \href{https://doi.org/10.48550/arXiv.2602.15758}{Feb. 2026}.
  \href{https://doi.org/10.48550/arXiv.2602.15758}
{doi: \textsf{%
10\hspace{.1pt}\discretionary{.}{%
}{.}\hspace{.4pt}48550\discretionary{/}{%
}{/}arXiv\hspace{.1pt}\discretionary{.}{%
}{.}\hspace{.4pt}2602\hspace{.1pt}\discretionary{.}{%
}{.}\hspace{.4pt}15758}}


\bibitem{lauDesignSpaceComputational2020}
\href{https://doi.org/10.1109/VL/HCC50065.2020.9127201}{S.~Lau, I.~Drosos,
  J.~M. Markel, and P.~J. Guo}.
\newblock \href{https://doi.org/10.1109/VL/HCC50065.2020.9127201}{{ {The
  {Design} {Space} of {Computational} {Notebooks}: {An} {Analysis} of 60
  {Systems} in {Academia} and {Industry}} }}.
\newblock \href{https://doi.org/10.1109/VL/HCC50065.2020.9127201}{In {\em 2020
  {IEEE} {Symposium} on {Visual} {Languages} and {Human}-{Centric} {Computing}
  ({VL}/{HCC})}}, \href{https://doi.org/10.1109/VL/HCC50065.2020.9127201}{pp.
  1--11}, \href{https://doi.org/10.1109/VL/HCC50065.2020.9127201}{Aug. 2020}.
  \href{https://doi.org/10.1109/VL/HCC50065.2020.9127201}
{doi: \textsf{%
10\hspace{.1pt}\discretionary{.}{%
}{.}\hspace{.4pt}1109\discretionary{/}{%
}{/}VL\discretionary{/}{%
}{/}HCC50065\hspace{.1pt}\discretionary{.}{%
}{.}\hspace{.4pt}2020\hspace{.1pt}\discretionary{.}{%
}{.}\hspace{.4pt}9127201}}


\bibitem{liuLostMiddleHow2024}
\href{https://doi.org/10.1162/tacl_a_00638}{N.~F. Liu, K.~Lin, J.~Hewitt,
  A.~Paranjape, M.~Bevilacqua, F.~Petroni, and P.~Liang}.
\newblock \href{https://doi.org/10.1162/tacl_a_00638}{{Lost in the {Middle}:
  {How} {Language} {Models} {Use} {Long} {Contexts}}}.
\newblock \href{https://doi.org/10.1162/tacl_a_00638}{{\em Transactions of the
  Association for Computational Linguistics}},
  \href{https://doi.org/10.1162/tacl_a_00638}{12:157--173},
  \href{https://doi.org/10.1162/tacl_a_00638}{2024}.
  \href{https://doi.org/10.1162/tacl_a_00638}
{doi: \textsf{%
10\hspace{.1pt}\discretionary{.}{%
}{.}\hspace{.4pt}1162\discretionary{/}{%
}{/}tacl\_a\_00638}}


\bibitem{marimoMarimoReactivePython2023}
\href{https://marimo.io/}{{marimo}}.
\newblock \href{https://marimo.io/}{{marimo: a reactive {Python} notebook}},
  \href{https://marimo.io/}{2023}.

\bibitem{marimoCollaborateAgentsUsing2026}
\href{https://docs.marimo.io/guides/generate_with_ai/marimo_pair/}{{marimo}}.
\newblock
  \href{https://docs.marimo.io/guides/generate_with_ai/marimo_pair/}{{Collaborate
  with agents using marimo pair}},
  \href{https://docs.marimo.io/guides/generate_with_ai/marimo_pair/}{June
  2026}.

\bibitem{marimoReactivityDocumentation2026}
\href{https://docs.marimo.io/guides/reactivity/}{{marimo}}.
\newblock \href{https://docs.marimo.io/guides/reactivity/}{{Understanding
  reactivity}}, \href{https://docs.marimo.io/guides/reactivity/}{2026}.

\bibitem{massonDirectGPTDirectManipulation2024}
\href{https://doi.org/10.1145/3613904.3642462}{D.~Masson, S.~Malacria,
  G.~Casiez, and D.~Vogel}.
\newblock \href{https://doi.org/10.1145/3613904.3642462}{{ {{DirectGPT}: {A}
  {Direct} {Manipulation} {Interface} to {Interact} with {Large} {Language}
  {Models}} }}.
\newblock \href{https://doi.org/10.1145/3613904.3642462}{In {\em Proceedings of
  the {CHI} {Conference} on {Human} {Factors} in {Computing} {Systems}}},
  \href{https://doi.org/10.1145/3613904.3642462}{May 2024}.
  \href{https://doi.org/10.1145/3613904.3642462}
{doi: \textsf{%
10\hspace{.1pt}\discretionary{.}{%
}{.}\hspace{.4pt}1145\discretionary{/}{%
}{/}3613904\hspace{.1pt}\discretionary{.}{%
}{.}\hspace{.4pt}3642462}}


\bibitem{mcnuttDesignAIpoweredCode2023a}
\href{https://doi.org/10.1145/3544548.3580940}{A.~M. Mcnutt, C.~Wang, R.~A.
  Deline, and S.~M. Drucker}.
\newblock \href{https://doi.org/10.1145/3544548.3580940}{{On the {Design} of
  {AI}-powered {Code} {Assistants} for {Notebooks}}}.
\newblock \href{https://doi.org/10.1145/3544548.3580940}{In {\em Proceedings of
  the 2023 {CHI} {Conference} on {Human} {Factors} in {Computing} {Systems}}},
  \href{https://doi.org/10.1145/3544548.3580940}{Apr. 2023}.
  \href{https://doi.org/10.1145/3544548.3580940}
{doi: \textsf{%
10\hspace{.1pt}\discretionary{.}{%
}{.}\hspace{.4pt}1145\discretionary{/}{%
}{/}3544548\hspace{.1pt}\discretionary{.}{%
}{.}\hspace{.4pt}3580940}}


\bibitem{monadjemiScopingReviewMixed2026}
\href{https://doi.org/10.1111/cgf.70434}{S.~Monadjemi, Y.~Guo, K.~Xu,
  A.~Endert, and A.~Crisan}.
\newblock \href{https://doi.org/10.1111/cgf.70434}{{ {A {Scoping} {Review} of
  {Mixed} {Initiative} {Visual} {Analytics} in the {Automation} {Renaissance}}
  }}.
\newblock \href{https://doi.org/10.1111/cgf.70434}{{\em Computer Graphics
  Forum}}, \href{https://doi.org/10.1111/cgf.70434}{p. e70434},
  \href{https://doi.org/10.1111/cgf.70434}{June 2026}.
  \href{https://doi.org/10.1111/cgf.70434}
{doi: \textsf{%
10\hspace{.1pt}\discretionary{.}{%
}{.}\hspace{.4pt}1111\discretionary{/}{%
}{/}cgf\hspace{.1pt}\discretionary{.}{%
}{.}\hspace{.4pt}70434}}


\bibitem{nationalgalleryofartNationalGalleryArt2021}
\href{https://github.com/NationalGalleryOfArt/opendata}{{National Gallery of
  Art}}.
\newblock \href{https://github.com/NationalGalleryOfArt/opendata}{{National
  {Gallery} of {Art} {Open} {Data} {Program}}},
  \href{https://github.com/NationalGalleryOfArt/opendata}{Apr. 2021}.

\bibitem{nationalgalleryofartOpenDataCommit2026}
\href{https://github.com/NationalGalleryOfArt/opendata/commit/
  e19fc9a6bf8167630be458f745dfff915fbe06ba}{{National Gallery of Art}}.
\newblock \href{https://github.com/NationalGalleryOfArt/opendata/commit/
  e19fc9a6bf8167630be458f745dfff915fbe06ba}{{National {Gallery} of {Art} {Open}
  {Data}: {August} 3, 2026 export}}.
\newblock \href{https://github.com/NationalGalleryOfArt/opendata/commit/
  e19fc9a6bf8167630be458f745dfff915fbe06ba}{Git commit \texttt{e19fc9a}},
  \href{https://github.com/NationalGalleryOfArt/opendata/commit/
  e19fc9a6bf8167630be458f745dfff915fbe06ba}{Aug. 2026}.

\bibitem{niuChartREGBenchmarkingImproving2026a}
\href{https://doi.org/10.48550/arXiv.2605.07415}{T.~Niu, Z.~Han, X.~Dong,
  Q.~Zhu, and W.~Che}.
\newblock \href{https://doi.org/10.48550/arXiv.2605.07415}{{ {{ChartREG}++:
  {Towards} {Benchmarking} and {Improving} {Chart} {Referring} {Expression}
  {Grounding} under {Diverse} referring clues and {Multi}-{Target} {Referring}}
  }}, \href{https://doi.org/10.48550/arXiv.2605.07415}{June 2026}.
  \href{https://doi.org/10.48550/arXiv.2605.07415}
{doi: \textsf{%
10\hspace{.1pt}\discretionary{.}{%
}{.}\hspace{.4pt}48550\discretionary{/}{%
}{/}arXiv\hspace{.1pt}\discretionary{.}{%
}{.}\hspace{.4pt}2605\hspace{.1pt}\discretionary{.}{%
}{.}\hspace{.4pt}07415}}


\bibitem{parkDirectVisEditingCodeBased2026a}
\href{https://doi.org/10.1109/PacificVis68791.2026.00014}{J.~Park, M.~An,
  H.~Yang, J.~Hwangbo, M.~H. Kim, H.~Jeon, and J.~Seo}.
\newblock \href{https://doi.org/10.1109/PacificVis68791.2026.00014}{{
  {{DirectVis}: {Editing} {Code}-{Based} {Interactive} {Visualization} with
  {Direct} {Manipulation}} }}.
\newblock \href{https://doi.org/10.1109/PacificVis68791.2026.00014}{In {\em
  2026 {IEEE} 19th {Pacific} {Visualization} {Conference} ({PacificVis})}},
  \href{https://doi.org/10.1109/PacificVis68791.2026.00014}{pp. 79--84}.
  \href{https://doi.org/10.1109/PacificVis68791.2026.00014}{IEEE},
  \href{https://doi.org/10.1109/PacificVis68791.2026.00014}{Sydney, Australia},
  \href{https://doi.org/10.1109/PacificVis68791.2026.00014}{Apr. 2026}.
  \href{https://doi.org/10.1109/PacificVis68791.2026.00014}
{doi: \textsf{%
10\hspace{.1pt}\discretionary{.}{%
}{.}\hspace{.4pt}1109\discretionary{/}{%
}{/}PacificVis68791\hspace{.1pt}\discretionary{.}{%
}{.}\hspace{.4pt}2026\hspace{.1pt}\discretionary{.}{%
}{.}\hspace{.4pt}00014}}


\bibitem{payandehNoteExInteractiveVisual2025}
\href{https://doi.org/10.48550/arXiv.2511.07223}{M.~H. Payandeh, L.-P. Yuan,
  and J.~Zhao}.
\newblock \href{https://doi.org/10.48550/arXiv.2511.07223}{{ {{NoteEx}:
  {Interactive} {Visual} {Context} {Manipulation} for {LLM}-{Assisted}
  {Exploratory} {Data} {Analysis} in {Computational} {Notebooks}} }},
  \href{https://doi.org/10.48550/arXiv.2511.07223}{Nov. 2025}.
  \href{https://doi.org/10.48550/arXiv.2511.07223}
{doi: \textsf{%
10\hspace{.1pt}\discretionary{.}{%
}{.}\hspace{.4pt}48550\discretionary{/}{%
}{/}arXiv\hspace{.1pt}\discretionary{.}{%
}{.}\hspace{.4pt}2511\hspace{.1pt}\discretionary{.}{%
}{.}\hspace{.4pt}07223}}


\bibitem{stahleVACPVisualAnalytics2026}
\href{https://doi.org/10.48550/arXiv.2603.29322}{T.~St\"{a}hle, P.~F. Gyarmati,
  T.~Spinner, R.~Sevastjanova, D.~Moritz, and M.~El-Assady}.
\newblock \href{https://doi.org/10.48550/arXiv.2603.29322}{{{VACP}: {Visual}
  {Analytics} {Context} {Protocol}}}.
\newblock \href{https://doi.org/10.48550/arXiv.2603.29322}{Accepted to IEEE VIS
  2026}, \href{https://doi.org/10.48550/arXiv.2603.29322}{Mar. 2026}.
  \href{https://doi.org/10.48550/arXiv.2603.29322}
{doi: \textsf{%
10\hspace{.1pt}\discretionary{.}{%
}{.}\hspace{.4pt}48550\discretionary{/}{%
}{/}arXiv\hspace{.1pt}\discretionary{.}{%
}{.}\hspace{.4pt}2603\hspace{.1pt}\discretionary{.}{%
}{.}\hspace{.4pt}29322}}


\bibitem{tangChartsCodeHierarchical2026}
\href{https://doi.org/10.18653/v1/2026.acl-long.616}{J.~Tang, H.~H. Zhao,
  L.~Wu, Z.~Zhang, Y.~Tao, D.~Mao, Y.~Wan, J.~Tan, M.~Zeng, M.~Li, and A.~J.
  Wang}.
\newblock \href{https://doi.org/10.18653/v1/2026.acl-long.616}{{ {From {Charts}
  to {Code}: {A} {Hierarchical} {Benchmark} for {Multimodal} {Models}} }}.
\newblock \href{https://doi.org/10.18653/v1/2026.acl-long.616}{In M.~Liakata,
  V.~P. Moreira, J.~Zhang, and D.~Jurgens, eds., {\em Proceedings of the 64th
  {Annual} {Meeting} of the {Association} for {Computational} {Linguistics}
  ({Volume} 1: {Long} {Papers})}},
  \href{https://doi.org/10.18653/v1/2026.acl-long.616}{pp. 13467--13566}.
  \href{https://doi.org/10.18653/v1/2026.acl-long.616}{Association for
  Computational Linguistics},
  \href{https://doi.org/10.18653/v1/2026.acl-long.616}{San Diego, California,
  United States}, \href{https://doi.org/10.18653/v1/2026.acl-long.616}{July
  2026}. \href{https://doi.org/10.18653/v1/2026.acl-long.616}
{doi: \textsf{%
10\hspace{.1pt}\discretionary{.}{%
}{.}\hspace{.4pt}18653\discretionary{/}{%
}{/}v1\discretionary{/}{%
}{/}2026\hspace{.1pt}\discretionary{.}{%
}{.}\hspace{.4pt}acl\discretionary{%
}{-}{-}long\hspace{.1pt}\discretionary{.}{%
}{.}\hspace{.4pt}616}}


\bibitem{wuB22020}
\href{https://doi.org/10.1145/3379337.3415851}{Y.~Wu, J.~M. Hellerstein, and
  A.~Satyanarayan}.
\newblock \href{https://doi.org/10.1145/3379337.3415851}{{ {B2: {Bridging}
  {Code} and {Interactive} {Visualization} in {Computational} {Notebooks}} }}.
\newblock \href{https://doi.org/10.1145/3379337.3415851}{In {\em Proceedings of
  the 33rd {Annual} {ACM} {Symposium} on {User} {Interface} {Software} and
  {Technology}}}, \href{https://doi.org/10.1145/3379337.3415851}{pp. 152--165}.
  \href{https://doi.org/10.1145/3379337.3415851}{ACM},
  \href{https://doi.org/10.1145/3379337.3415851}{Virtual Event USA},
  \href{https://doi.org/10.1145/3379337.3415851}{Oct. 2020}.
  \href{https://doi.org/10.1145/3379337.3415851}
{doi: \textsf{%
10\hspace{.1pt}\discretionary{.}{%
}{.}\hspace{.4pt}1145\discretionary{/}{%
}{/}3379337\hspace{.1pt}\discretionary{.}{%
}{.}\hspace{.4pt}3415851}}


\bibitem{wuStepMINDVisualFramework2026}
\href{https://doi.org/10.1145/3742413.3789070}{Y.~Wu, Y.~Wan, M.~El-Assady, and
  A.~Y. Wang}.
\newblock \href{https://doi.org/10.1145/3742413.3789070}{{ {{StepMIND}: {A}
  {Visual} {Framework} for {Stepwise}, {Multimodal}, and {Bidirectional}
  {Explanations} of {AI}-{Generated} {Data} {Analysis} {Pipeline}} }}.
\newblock \href{https://doi.org/10.1145/3742413.3789070}{In {\em Proceedings of
  the 31st {International} {Conference} on {Intelligent} {User} {Interfaces}}},
  \href{https://doi.org/10.1145/3742413.3789070}{{IUI} '26},
  \href{https://doi.org/10.1145/3742413.3789070}{pp. 522--539}.
  \href{https://doi.org/10.1145/3742413.3789070}{Association for Computing
  Machinery}, \href{https://doi.org/10.1145/3742413.3789070}{New York, NY,
  USA}, \href{https://doi.org/10.1145/3742413.3789070}{Mar. 2026}.
  \href{https://doi.org/10.1145/3742413.3789070}
{doi: \textsf{%
10\hspace{.1pt}\discretionary{.}{%
}{.}\hspace{.4pt}1145\discretionary{/}{%
}{/}3742413\hspace{.1pt}\discretionary{.}{%
}{.}\hspace{.4pt}3789070}}


\bibitem{WaitGPTMonitoringSteering2024}
\href{https://doi.org/10.1145/3654777.3676374}{L.~Xie, C.~Zheng, H.~Xia, H.~Qu,
  and Z.-T. Chen}.
\newblock \href{https://doi.org/10.1145/3654777.3676374}{{ {{WaitGPT}:
  {Monitoring} and {Steering} {Conversational} {LLM} {Agent} in {Data}
  {Analysis} with {On}-the-{Fly} {Code} {Visualization}} }}.
\newblock \href{https://doi.org/10.1145/3654777.3676374}{In {\em Proceedings of
  the 37th {Annual} {ACM} {Symposium} on {User} {Interface} {Software} and
  {Technology}}}, \href{https://doi.org/10.1145/3654777.3676374}{p. Article
  119}. \href{https://doi.org/10.1145/3654777.3676374}{Association for
  Computing Machinery},
  \href{https://doi.org/10.1145/3654777.3676374}{Pittsburgh, PA, USA},
  \href{https://doi.org/10.1145/3654777.3676374}{2024}.
  \href{https://doi.org/10.1145/3654777.3676374}
{doi: \textsf{%
10\hspace{.1pt}\discretionary{.}{%
}{.}\hspace{.4pt}1145\discretionary{/}{%
}{/}3654777\hspace{.1pt}\discretionary{.}{%
}{.}\hspace{.4pt}3676374}}


\bibitem{zhaoSupportingHandoffAsynchronous2018}
\href{https://doi.org/10.1109/TVCG.2017.2745279}{J.~Zhao, M.~Glueck,
  P.~Isenberg, F.~Chevalier, and A.~Khan}.
\newblock \href{https://doi.org/10.1109/TVCG.2017.2745279}{{ {Supporting
  {Handoff} in {Asynchronous} {Collaborative} {Sensemaking} {Using}
  {Knowledge}-{Transfer} {Graphs}} }}.
\newblock \href{https://doi.org/10.1109/TVCG.2017.2745279}{{\em IEEE
  Transactions on Visualization and Computer Graphics}},
  \href{https://doi.org/10.1109/TVCG.2017.2745279}{24(1):340--350},
  \href{https://doi.org/10.1109/TVCG.2017.2745279}{Jan. 2018}.
  \href{https://doi.org/10.1109/TVCG.2017.2745279}
{doi: \textsf{%
10\hspace{.1pt}\discretionary{.}{%
}{.}\hspace{.4pt}1109\discretionary{/}{%
}{/}TVCG\hspace{.1pt}\discretionary{.}{%
}{.}\hspace{.4pt}2017\hspace{.1pt}\discretionary{.}{%
}{.}\hspace{.4pt}2745279}}


\bibitem{zhaoChartEditHowFar2025}
\href{https://doi.org/10.18653/v1/2025.findings-acl.185}{X.~Zhao, X.~Liu,
  Y.~Haoyue, X.~Luo, F.~Zeng, J.~Li, Q.~Shi, and C.~Chen}.
\newblock \href{https://doi.org/10.18653/v1/2025.findings-acl.185}{{
  {{ChartEdit}: {How} {Far} {Are} {MLLMs} {From} {Automating} {Chart}
  {Analysis}? {Evaluating} {MLLMs}' {Capability} via {Chart} {Editing}} }}.
\newblock \href{https://doi.org/10.18653/v1/2025.findings-acl.185}{In W.~Che,
  J.~Nabende, E.~Shutova, and M.~T. Pilehvar, eds., {\em Findings of the
  {Association} for {Computational} {Linguistics}: {ACL} 2025}},
  \href{https://doi.org/10.18653/v1/2025.findings-acl.185}{pp. 3616--3630}.
  \href{https://doi.org/10.18653/v1/2025.findings-acl.185}{Association for
  Computational Linguistics},
  \href{https://doi.org/10.18653/v1/2025.findings-acl.185}{Vienna, Austria},
  \href{https://doi.org/10.18653/v1/2025.findings-acl.185}{July 2025}.
  \href{https://doi.org/10.18653/v1/2025.findings-acl.185}
{doi: \textsf{%
10\hspace{.1pt}\discretionary{.}{%
}{.}\hspace{.4pt}18653\discretionary{/}{%
}{/}v1\discretionary{/}{%
}{/}2025\hspace{.1pt}\discretionary{.}{%
}{.}\hspace{.4pt}findings\discretionary{%
}{-}{-}acl\hspace{.1pt}\discretionary{.}{%
}{.}\hspace{.4pt}185}}


\end{thebibliography}
\endgroup

\end{document}